\documentclass{elsart}

\usepackage{graphicx}

\usepackage{amssymb,amsmath}

\begin{document}

\begin{frontmatter}


\title{Evolution of gravitational waves from inflationary brane-world :
        \\ numerical study of high-energy effects}
%
\author[phys]{Takashi Hiramatsu},
\ead{hiramatsu@utap.phys.s.u-tokyo.ac.jp}
\author[phys]{Kazuya Koyama},
\ead{kazuya@utap.phys.s.u-tokyo.ac.jp}
\author[phys,RES]{Atsushi Taruya}
\ead{ataruya@utap.phys.s.u-tokyo.ac.jp}
\address[phys]{Department of physics ,The University of Tokyo, Tokyo 113-0033, Japan}
\address[RES]{Research Center for the Early Universe, The University of Tokyo, Tokyo 113-0033, Japan}
%
%
%
%
%
\begin{abstract}
We study the evolution of gravitational waves(GWs) after 
inflation in a brane-world cosmology embedded in five-dimensional 
anti-de Sitter spacetime. 
Contrary to the standard four-dimensional results, the GWs 
at the high-energy regime in brane-world model suffer from the 
effects of the non-standard cosmological expansion and the excitation of the 
Kaluza-Klein modes(KK-modes), which can affect the amplitude of stochastic 
gravitational wave background significantly. 
To investigate these two high-energy effects quantitatively, we numerically 
solve the wave equation of the GWs in the radiation dominated epoch
at relatively low-energy scales. 
We show that the resultant GWs are suppressed by the 
excitation of the KK modes. The created KK modes are rather soft 
and escape away from the brane to the bulk gravitational field. 
The results are also compared to the semi-analytic prediction 
from the low-energy approximation and the evolved amplitude of GWs on 
the brane reasonably matches the numerical simulations. 
\end{abstract}
%
%
%
%
%
%
%
%
%
%
\end{frontmatter}
%
%

\section{Introduction}
\label{sec: intro}

The stochastic gravitational wave background 
(stochastic GWB) generated during the inflationary era is 
a promising source for the low-frequency gravitational waves  
and this can provide a direct way to probe the extremely early Universe. 
In particular, such GWs are expected to have information 
about the extra dimensions. Inspired by the recent development in 
particle physics, there has been a lot of 
active debate on the possibility that we live in a brane which is a 
three dimensional world embedded in a higher dimensional space 
\cite{RS1}\cite{RS2}\cite{Bin}\cite{Lan}. If this is true, the 
standard prediction of stochastic GWB from the four-dimensional theory 
is dramatically altered and the GWB can be a powerful tool to 
discriminate the presence/absence of the extra-dimension. 
For instance, in the case of the single-brane model proposed by 
Randall \& Sundrum\cite{RS2}, the scale of the extra 
dimension is well below the length $l\lesssim0.1$mm according to the 
current experiments of the Newton gravity (e.g., \cite{Chi}). 
Therefore, the effect of the extra-dimension can be imprinted on the GWB  
at the low-frequency band, $f\gtrsim f_{\rm crit}\sim 2 \times 10^{-4}\,(0.1\mbox{mm}/l)^{1/2}$ Hz 
\cite{Hog}\cite{Mag}.

Despite the above interesting suggestions, however, quantitative 
prediction for the stochastic GWB in the brane-world cosmology is still under 
investigation. The spectrum of GWs generated by de Sitter inflation on 
a brane was first calculated by \cite{Lan}. And later, the cosmological 
evolution of the GWs has been analytically studied in a very idealistic situation:  
the transition from the de-Sitter brane to the Minkowski brane \cite{Gor} 
and the transition from the de-Sitter brane to the de-Sitter brane with 
different cosmological constant \cite{Koba}. For more realistic situation 
of the cosmological evolution with equation of state $p=w\rho$, however, 
the wave equation of GWs cannot be separable and the analytical treatment 
is generally intractable. We must tackle the complicated partial differential 
equation directly.

Theoretically, the evolution of GWs in the brane-world cosmology is expected to 
deviate from the standard four-dimensional theory in the following two aspects: 
1) the non-standard cosmological expansion due to the bulk gravity 
and 2) the excitation of the Kaluza-Klein mode(KK-mode) of graviton. The former 
may enhance the amplitude of GWs and the latter may suppress or modulate the  
GW form on a brane. These effects are particularly significant in the 
high-energy regime of the universe. An important question is which effect is 
dominant and how the amplitude of GWs changes in a realistic situation of the 
cosmological brane-world.

For this purpose, we solve the wave equation of GWs numerically 
in the Randall-Sundrum type single-brane model. It has been shown that the excitation 
of KK-modes are suppressed during inflation and zero-mode remains constant after 
inflation at super-horizon scales \cite{Lan}. Thus we will consider the evolution of 
the GWs starting from an initially zero-mode state at super-horizon scales 
and focus on the behavior just after the horizon-crossing time. 
In the high-energy regime, the separation between 
zero-mode and KK-modes does not hold for the modes with wavelength shorter than 
the horizon. This implies that even in an initially zero-mode state, 
KK-modes are inevitably excited when the perturbation crosses the horizon. 
Thus, we can observe how the excitation of the KK modes modifies the behavior 
of the perturbations.

We set up the basic equations 
in section \ref{sec: basic_equation}. Based on these equations, the 
numerical calculation of wave equation is performed and the results are 
presented in section \ref{sec: numerical_analysis}. To understand the 
behavior on the brane, section \ref{sec: low-energy_expansion} 
describes the semi-analytic treatment. Using the low-energy approximation, 
we derive an effective equation of the GWs on the brane. The resultant equation
reduces to the ordinary differential equation and the numerical results of 
this equation reasonably match those of the full numerical treatment. 
Final section \ref{sec: summary} is devoted to the summary and discussion.

\section{Basic equations}
\label{sec: basic_equation}

In this letter, we specifically treat the single-brane model 
embedded in a five-dimensional anti-de Sitter space, in which the matter content 
on brane is simply given by a homogeneous and isotropic perfect fluid satisfying the 
equation of state $p=w\rho$. Using the Gaussian normal coordinate to the brane, 
the background metric is given by    
\begin{equation}
ds^2 = -n^2(t,y)dt^2 + a^2(t,y)\delta_{ij}dx^idx^j+dy^2, 
\label{eq:metric}
\end{equation}
where the brane is located at $y=0$. 
The lapse function $n(t,y)$ and the scale factor $a(t,y)$ 
for the background space-time are determined from the five-dimensional 
Einstein equation. In absence of the dark radiation, these quantities are 
written as follows \cite{Bin}\cite{Lan}:
\begin{align}
 a(t,y) &= a_0(t)\left\{\cosh(\mu y) - 
\left(1+\frac{\rho(t)}{\lambda}\right)\sinh(\mu y)\right\}, \\
 \begin{split}
 n(t,y) &= \frac{\dot{a}(t,y)}{\dot{a}_0(t)} \\
        &= e^{-\mu y}+(2+3w)\frac{\rho(t)}{\lambda}\sinh(\mu y),
 \end{split}
\end{align}
where $\lambda>0$ denotes the tension of the brane and $\mu^{-1}$ 
represents the curvature scale of the anti-de Sitter bulk, 
which is related to the tension $\lambda$ by $\mu = \sqrt{\kappa_4^2\lambda/3}$. 
The scale factor $a_0(t)$ and the energy density on the brane $\rho(t)$ 
are determined from the effective Friedmann equations \cite{Bin}:
\begin{equation}
 H^2\equiv \left(\frac{\dot{a}_0}{a_0}\right)^2 = 
\frac{\kappa_4^2}{3}\rho(t)\left(1+\frac{\rho(t)}{2\lambda}\right),\qquad 
\dot{\rho}=-3H(1+w)\rho(t),
\label{eq:Hubble}
\end{equation}
The solution of the above equations is easily obtained and 
can be expressed in terms of the dimensionless variables $\tau= t/t_*$ and 
$\epsilon=\rho/\lambda$:  
\begin{align}
 a_0(\tau)&= a_*\left(\tau^2 + 2c\tau - 2c\right)^{\frac{1}{3(1+w)}}, \\
 \rho(\tau)&= \rho_*\left(\tau^2 + 2c\tau - 2c\right)^{-1}, \\
 c &\equiv \sqrt{1+\frac{\epsilon_*}{2}}-1, 
\end{align}
where the subscript $*$ means the quantity evaluated at the time 
$\tau=1$. Note that the Gaussian-normal coordinate has a coordinate 
singularity at $y=y_{\text{c}}$, defined by $a(\tau,y_{\text{c}}(\tau))=0$:
\begin{equation}        
 \mu y_{\text{c}}(\tau) = \frac{1}{2}\log\left\{\frac{2}{\epsilon_*}\left(\tau+c\right)^2\right\}.
\label{eq:singularity}  
\end{equation}

Now consider the tensor perturbation in the metric 
(\ref{eq:metric}). The perturbed metric becomes 
\begin{equation}
ds^2 = -n^2(t,y)dt^2 + a^2(t,y)(\delta_{ij}+E_{ij})dx^idx^j+dy^2. 
\end{equation}
The perturbed quantity $E_{ij}$ satisfies the transverse and the traceless 
conditions, which is automatically gauge-invariant. 
Decomposing the perturbation in spatial Fourier modes as 
$E_{ij}=E(t,y;k) e^{ikx}\hat{e}_{ij}$, where $\hat{e}_{ij}$ is 
transverse-traceless polarization tensor, 
the wave equation for the Fourier component $E$ becomes \cite{Lan}: 
\begin{equation}
\frac{\partial^2E}{\partial t^2}+\left(3\frac{\dot{a}}{a}-\frac{\dot{n}}{n}\right)\frac{\partial E}{\partial t}+\frac{n^2}{a^2}k^2E-n^2\left\{\frac{\partial^2E}{\partial y^2}+\left(3\frac{a'}{a}+\frac{n'}{n}\right)\frac{\partial E}{\partial y}\right\} = 0, 
\label{eq:evolution}
\end{equation}
where the dot and the prime respectively denote the derivative with respect to 
the time $t$ and the bulk coordinate $y$. The above equation must be solved 
under the boundary condition at the brane, 
which follows from the perturbed Israel condition. 
Ignoring the anisotropic stress tensor $\pi_{ij}$ arising from the matter on 
brane, we obtain the junction condition:
\begin{equation}
 \left.\frac{\partial E}{\partial y}\right|_{y=0}=0.
\end{equation}

Hereafter, we will focus on the time evolution of the tensor perturbation 
in the radiation-dominated epoch($w=1/3$) and solve the wave equation 
(\ref{eq:evolution}) numerically.

\section{Numerical analysis}
\label{sec: numerical_analysis}

To solve the wave equation (\ref{eq:evolution}) numerically,
we use the spectral method, which is a standard technique
in the computational fluid dynamics \cite{Can} and recently becomes popular in 
subject of numerical relativity (e.g., \cite{Bon}). 
To be precise, we adopt a Tchebychev collocation method
with the Gauss-Lobatto collocation points,  
$y_\ell=\cos(\ell\pi/N)$ with $\ell=0, 1, \cdots,N$.
With this method, the quantity $E(\tau,y)$ is first transformed to a set of variables 
$E_{\ell}(\tau)$ defined in the Tchebychev space, and the 
partial differential equation (\ref{eq:evolution}) 
can be regarded as the ordinary differential equations for $E_{\ell}$. 
Then, the time evolution of $E_{\ell}$ is followed by the Predictor-Corrector 
method based on the Adams-Bashforth-Moulton finite-difference scheme.

As mentioned in section \ref{sec: intro}, 
the initial condition for the quantity $E$ is set to the 
zero-mode solution in the inflationary era, i.e., $E(\tau_0,y)=$ constant. 
The calculations are then started before the wavelength of the zero-mode 
crosses the Hubble horizon. For convenience, we set the comoving wave number 
$k$ to $k=a_*H_*$, that is, the GW mode just crosses the Hubble horizon 
when $\tau=1$.

The difficulty in present numerical calculation is that the computational domain 
should be finite. In addition, the Gaussian normal coordinate (\ref{eq:metric}) 
has a coordinate singularity (see Eq.(\ref{eq:singularity})). 
Hence, we must introduce another boundary in the bulk and impose another boundary 
condition at this boundary for numerical purpose. Perhaps, the simplest choice is 
the static boundary, however, due to the narrow computational domain, 
the suppression of light KK-modes becomes significant and the 
artifitial reflection wave affects the physical brane.  
In this letter, to remedy this, we put a moving 
regulator brane at $y_{\text{reg}}(\tau)=\gamma\, y_{\text{c}}(\tau)$ 
inside the coordinate singularity and impose the Neumann boundary 
condition\footnote{Restricting the parameter to $\epsilon_*<1$,  
the calculation with the static boundary can reproduce
the results using the moving boundary at an early time, 
however, the late-time pahse of the calculation is significantly affected 
by the artifitial reflection wave. }:
\begin{equation}
 \left.n^{\mu}\frac{\partial E}
{\partial x^\mu}\right|_{y=y_{\scriptscriptstyle\rm reg}(\tau)}=0,
\label{eq:Neumann}
\end{equation}
where $n^\mu$ is the normal vector to the boundary trajectory. 
To impose this boundary condition in the spectral method, we introduce 
the new coordinate $z=2y/y_{\text{reg}}(\tau)-1$, and solve (\ref{eq:evolution}) 
in the fixed range $z:[-1,1]$. Thus, provided the location of the boundaries  
$y=0$ and $y=y_{\text{reg}}(\tau)$, the only physical parameter is $\epsilon_*$, 
i.e., the dimensionless energy density $\rho/\lambda$ at the horizon-crossing time, 
which is directly related to the scale of the GW concerned. 
For large $\epsilon_*$, the physical brane is very close to 
the coordinate singularity and the computational domain of our code 
becomes narrow.  
This causes pathological behavior near the coordinate singularity.  
Hence, in this letter, we restrict our analysis to the 
the small $\epsilon_*$ case. 

\begin{figure}[!ht]
 \centering
 \includegraphics[width=10cm]{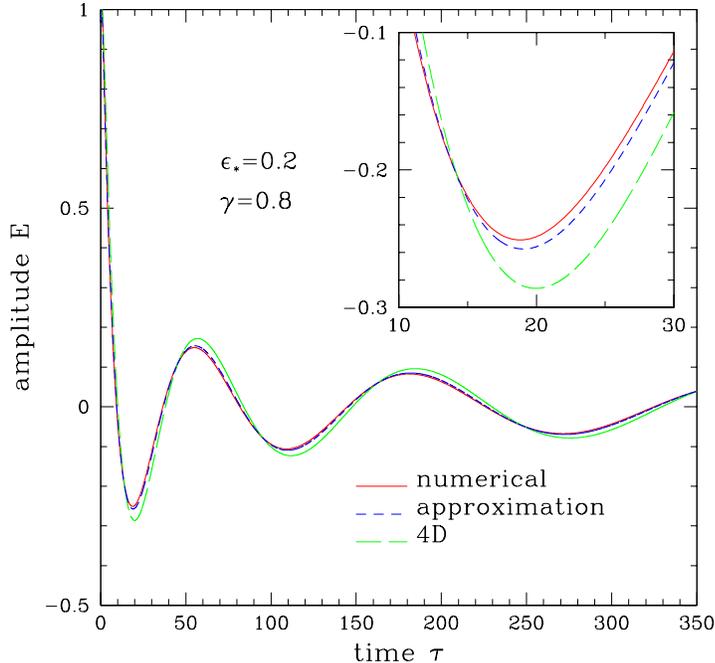}
 \caption{Numerical solution of GW on brane ({\it solid}), together with 
the 4-dimensional GW ({\it long-dashed}) and the result from the low-energy 
approximation ({\it short-dashed}, see Sec.\ref{sec: low-energy_expansion}). 
The upper right window is the zoom-up of the solutions in the range $10<\tau<30$.
 \label{fig: sol_on_brane}} 
\end{figure}
\begin{figure}[!ht]
 \centering
 \includegraphics[width=10cm]{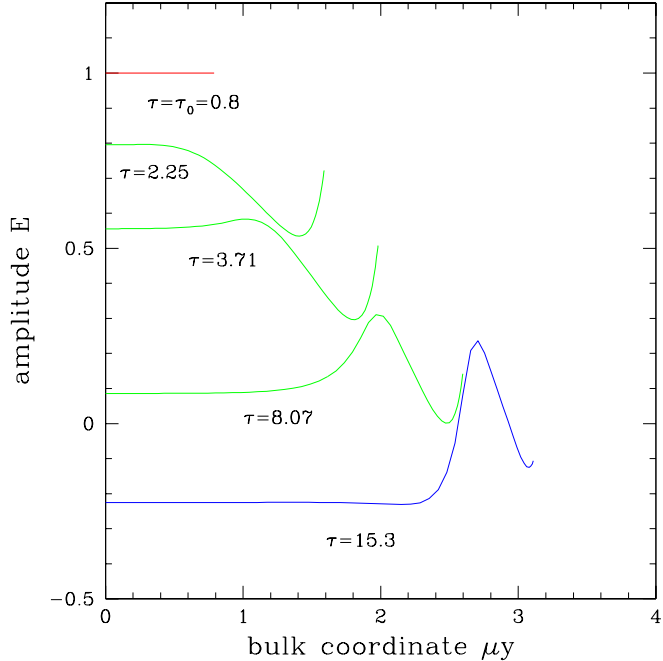}
 \caption{Snapshot of the waveform as function of the bulk coordinate $y$ 
        in the case with the parameters $(\epsilon_*,\gamma)=(0.2, 0.8)$. 
 \label{fig: snapshot}}
\end{figure}

In Fig.\ref{fig: sol_on_brane}, 
the resultant waveform on the brane is depicted as function of time 
in the case with $\epsilon_*=0.2$, $\gamma=0.8$ and $N=64$. 
Apart from the overall damping $1/a_0(t)$ due to the cosmological expansion, 
the amplitude of the numerical solution on the brane ({\it solid}) 
becomes smaller than that of the solution in the four-dimensional theory 
({\it long-dashed}). This result simply reflects the fact that the localization of 
gravity is not fulfilled in presence of the $\rho^2$-term (see Eq.(\ref{eq:Hubble})) 
and the GW can easily escape from the brane to the bulk. 

\begin{figure}[!ht]
 \centering
 \includegraphics[width=10cm]{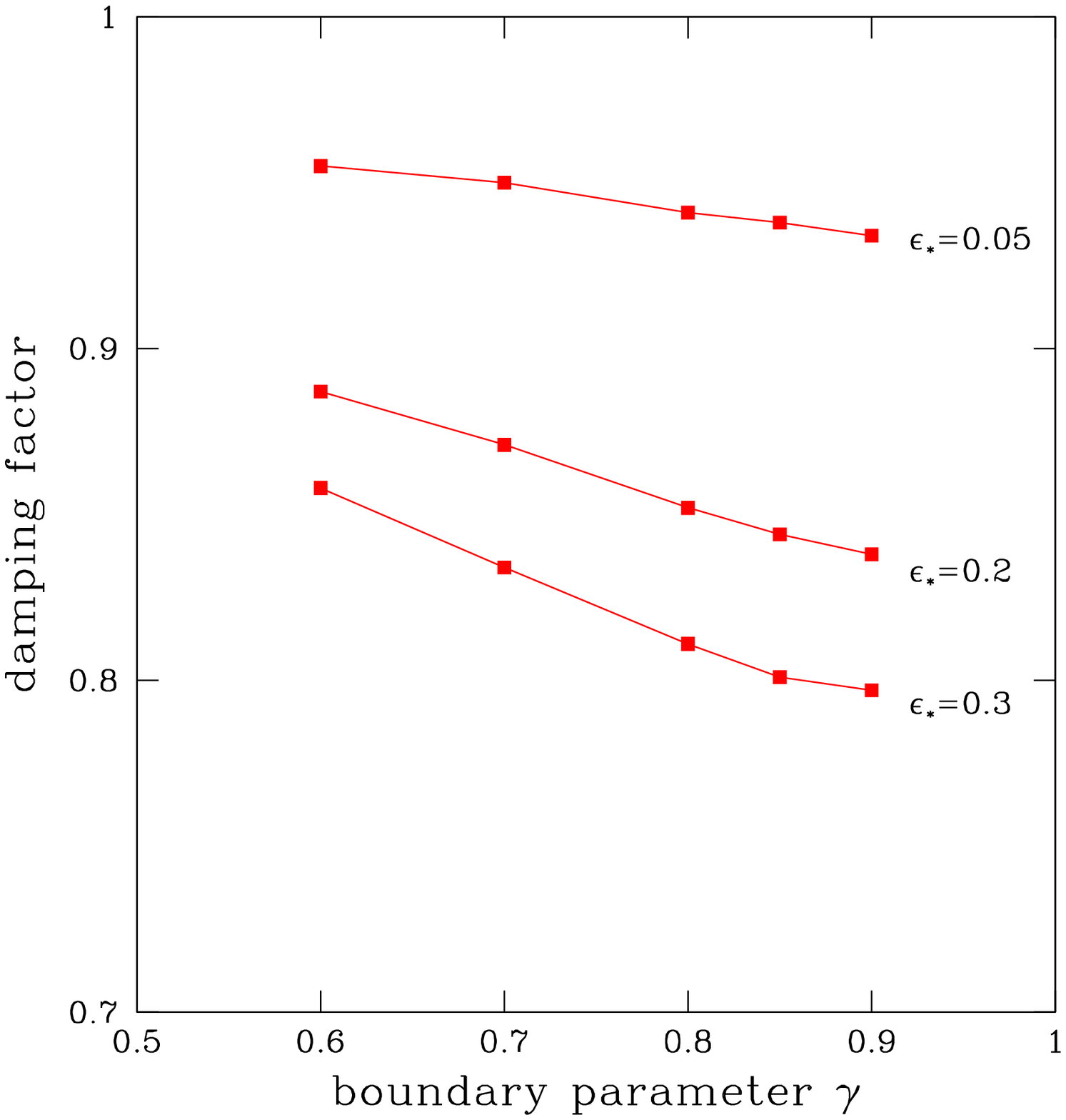}
 \caption{Damping factor of the GW amplitude on brane as function of 
        boundary parameter $\gamma$
 \label{fig: damping}}
\end{figure}

Fig.\ref{fig: snapshot} shows the snapshots of the waveform plotted as 
function of the original coordinate $y$. 
In the low-energy limit ($\rho/\lambda\ll1$) of the present case, 
the equation (\ref{eq:evolution}) becomes 
separable and the analytic solution can be obtained as $E(\tau,y)=u(y)\phi(\tau)$. 
The mode function for the bulk coordinate $u(y)$ is given by 
\begin{equation}
 u_0(y) = \text{constant}, 
\label{eq: zero-mode}
\end{equation}
for the zero-mode and 
\begin{equation}
 u_m(y) = e^{2\mu y}H^{(1)}_2\left(\frac{m}{\mu}e^{\mu y}\right), 
\label{eq: KK-mode}
\end{equation}
for the KK-mode with the effective mass $m$. The function  $H^{(1)}_2(z)$ 
denotes the Hankel function for the second kind. 
We attempted to approximate the wave form of the numerical result 
by the superposition of the 
    mode funcitons (\ref{eq: zero-mode}) and (\ref{eq: KK-mode}). 
    If we set the mass of  
    the KK-mode to the Hubble parameter at the horizon crossing time, 
    the mode functions roughly reproduce the waveform of the 
    numerical result, although the agreement is at a qualitative 
    level. It is clear from Figs. \ref{fig: sol_on_brane} and 
    \ref{fig: snapshot} that the created KK-modes is 
    rather light, consistent with our numerical set-up 
    with $\epsilon_*\lesssim1$, however, to obtain the KK-mode spectrum, 
    a more quantitative analysis is needed. 
 
The introduction of the regulator brane in the bulk is certainly not 
a proper way to treat the boundary condition in the bulk. 
The regulator brane may eventually affect the physical brane located at $y=0$. 
As shown in Fig.\ref{fig: sol_on_brane}, however, the 
excitation of the massive KK-modes becomes significant only 
at the horizon crossing time. The modification of the 
amplitude is almost determined around the horizon crossing and 
we need not follow the evolution of the perturbations for a long time. 
Thus, if the regulator brane is located sufficiently far away from the physical 
brane, we expect that we can safely neglect the effect of the 
regulator brane. To confirm this, in Fig.3, we plot the damping factor, i.e., 
the ratio of the amplitude obtained from the numerical solution
to that from the four-dimensional theory in the case of 
$\epsilon_*=0.05$, $0.2$ and $0.3$.
The number of collocation points used 
here is $N=64$ except for the cases of the boundary parameters 
$\gamma=0.85$ and $0.9$ with $N=128$. 
Since the suppression of GW amplitude is effective only when the mode crosses 
the Hubble horizon, the damping factor evaluated at the 
end of the calculaiton is time-independent. 
The damping factor depends linearly on small $\epsilon_*$, 
and tends to converge for $\gamma \rightarrow 1$.

\section{Result from low-energy expansion}
\label{sec: low-energy_expansion}

In order to understand the behavior of the numerical results in previous section, 
we employ a low energy expansion scheme to solve the wave equation approximately. 
In this treatment, the term $\rho/\lambda$ in effective 
Friedmann equation (\ref{eq:Hubble}) is assumed to be small. Thus, 
treating the dimensionless 
quantity $\epsilon=\rho/\lambda$ as small expansion parameter, 
we expand the tensor perturbation $E(\tau,y)$, the scale factor $a_0(\tau)$
and the Hubble parameter $H(\tau)$:  
\begin{align}
 E(\tau,y) &= E_0(\tau) + E_1(\tau,y) + E_2(\tau,y) + \cdots, 
\label{eq:Exp1} \\
 a_0(\tau) \equiv e^{\alpha(\tau)} &=
e^{\alpha_0(\tau)}(1+\alpha_1(\tau)+\cdots), 
\label{eq:Exp2} \\
 H(\tau) &= H_0(\tau) + H_1(\tau)+\cdots, \label{eq:Exp3}
\end{align}
 where $E_k(\tau,y)$, $\alpha_k(\tau)$ and $H_k(\tau)$ are the functions
of the order $\mathcal{O}(\epsilon^k)$. 
Note further the fact that in the low energy approximation, 
the time-derivative of the lower-order quantities is comparable to the 
$y$-derivative of the higher-order ones \cite{KK}. 
This is because the time derivative is of the order of 
the Hubble parameter, while the $y$-derivative is of the order of the bulk 
scale $\mu$. Thus, we have:
\begin{equation}
 \left(\frac{\dot{\alpha}}{\alpha'}\right)^2 \sim \mathcal{O}(\epsilon).
\label{eq: low_energy_expansion}
\end{equation}
Keeping the relation (\ref{eq: low_energy_expansion}) in mind, 
one can expand the wave equation (\ref{eq:evolution}) in terms of $\epsilon$ and 
find the higher-order solutions iteratively. Then, imposing boundary conditions, 
we derive an ordinary differential equations for $E_0(\tau)$ \cite{TH}.

The derivation of the equation for $E_0(\tau)$ is straightforward but 
lengthly calculation. The final form of the equation up to the first order 
in $\epsilon$ is 
\begin{equation}
 \begin{split}
  \Box_\tau E_0 &= 
\frac{9\epsilon_*(1+w)^2}{8}\left(1-2B(y_0)\right)\Box^2_\tau E_0 \\
  &-\frac{\rho}{\lambda}\left\{(2+3w) - (3w+5)B(y_0)\right\}\Box_\tau E_0 
\\
  &+\left\{3\frac{H_1}{H_0}-\frac{3}{2}(5+3w)(1+w)\frac{\rho}{\lambda}\left
(1-2B(y_0)\right) 
\right\}H_0\dot{E}_0 \\
  &+\left\{-2\alpha_1 - 3(1+w)\frac{\rho}{\lambda}\left(1-2B(y_0)\right) 
\right\}\left(\frac{k}{e^{\alpha_0}}\right)^2E_0,
 \end{split} \label{eq:LEE}
\end{equation}
in the case of the static boundary. Here, we defined 
the operator $\Box_{\tau }$ and the quantity $B$ as follows: 
\begin{equation}
 \Box_\tau E_0 \equiv -\,\frac{\partial^2 E_0}{\partial \tau^2}-3H_0\frac{\partial E_0}{\partial \tau} - \left(\frac{k}{e^{\alpha_0}}\right)^2E_0,
\end{equation}
\begin{equation}
 B(y_0) \equiv \frac{\mu y_0}{1-e^{-2\mu y_0}}.
\end{equation}
The variable  $y_0$ denotes the position of the boundary determined 
at an initial time $\tau=\tau_0$, i.e., $y_0=\gamma y_c(\tau_0)$.
If the regulator brane is moving, the equation contains the 
term proportional to $\dot{y}_{\rm reg}(t)$ and it becomes rather
complicated. Note also that this approximation will 
break down if the regulator brane becomes far away from our brane:
\begin{equation}
e^{\mu y_{\rm reg}} > \rho/\lambda,
\end{equation}
since the low energy approximation breaks down on the regulator brane.

In spite of the above limitations of this approximation, 
the numerical results for $E_0(\tau)$ quite well reproduces 
the full numerical simulation even in the case of 
%
the moving boundary\footnote{Note also that the numerical integration of 
effective equation (\ref{eq:LEE}) reproduces the full numerical 
simulation with the static boundary if $\epsilon_*<1$.}
%
(see {\it short-dashed} in Fig.\ref{fig: sol_on_brane}).
Thus, at least in a qualitative level, we can 
use this approximation to understand the behavior of perturbations on the 
brane.

The effective equation (\ref{eq:LEE}) contains a higher derivative term 
$\Box_\tau^2 E_0$ 
which describes the non-local effect caused by the propagation of the wave 
in the bulk.  If we numerically solve the equation with this higher-derivative  
term, we find that the solution badly diverges in time. It might be caused by  
the truncation of infinite numbers of higher derivative terms at a finite order.
Hence we neglect the $\Box_\tau^2 E_0$ term when we solve the ordinary  
differential equation (\ref{eq:LEE}) numerically. 
The first terms in the coefficients of $H_0\dot{E}_0$ and 
$k^2 e^{-2 \alpha_0}E_0$ 
come from the modification of background Friedmann equation, that is,  
$\rho^2$-term in Eq. (\ref{eq:Hubble}).  
The second terms arise from the non-separable nature of the metric 
(\ref{eq:metric}), which can be deduced from the fact that they vanish if we 
take $w=-1$.  These are the terms which excite KK-modes in the bulk 
and cause the dissipation of the perturbations on the brane.

In the long wavelength limit $k \to 0$, the zero-mode solution $E_0=$ const.  
is the solution for the effective equation (\ref{eq:LEE}).  
However, if the perturbation crosses the horizon, the zero-mode solution 
which satisfies $\Box_{\tau} E_0 =0$ cannot be a solution. Then the  
KK-modes are inevitably excited. The effects of the non-standard cosmological
expansion tend to enhance the amplitude of the perturbation compared to
the four-dimensional theory. On the other hand, both the numerical results 
and the low energy approximation show that the amplitude of the perturbations 
decreases, which implies that the influence of the excitation of KK-modes overcomes
the effects of the non-standard cosmological expansion. Therefore, 
the suppression of the amplitude of GWs can be understood as the consequence of 
the excitation of the KK-modes due to the non-separable nature of the bulk metric 
in the Gaussian normal coordinate defined with respect to the physical brane. 


\section{Summary and Discussion}
\label{sec: summary}

In this letter, we have numerically investigated the evolution of GWs 
during the radiation dominated epoch in the Randall-Sundrum type single-brane  
model. Especially focusing on the behavior after the horizon-crossing time, 
we found that the amplitude of GWs on brane is suppressed  
compared to that of the four-dimensional theory. 
To interpret the numerical results, we also employ 
the low energy approximation and derive the effective wave equation on 
the brane (Eq.(\ref{eq:LEE})). The solution of this equation 
reasonably agrees with numerical simulations and the 
suppression of GWs can be understood as a consequence of the excitation 
of KK-modes. Although the created KK-modes seem to be rather soft, 
the influence of the KK-modes on GWs overcomes the effect of the non-standard 
cosmological expansion arising from the $\rho^2$-term. 
The suppression of GWs becomes significant as increasing the energy scales 
$\epsilon_*$. Therefore, contrary to the four-dimensional prediction, 
the intensity of the stochastic GWB around the frequency $f\sim f_{\rm crit}$ 
tends to decrease as increasing the frequency, which might provide an important 
clue to probe the presence or the absence of extra-dimensions.

For more quantitative prediction for the spectrum of stochastic GWB, 
however, the present numerical analysis has several limitations. 
While the evolved amplitude of GWs is examined at the relatively low 
energy scales $\epsilon_*\lesssim1$, corresponding to the low-frequency modes 
with $f\lesssim f_{\rm crit}$, 
the high-frequency GWs with $f\gtrsim f_{\rm crit}$ are expected to 
suffer from the high-energy effects more seriously. As pointed out by 
\cite{Lan-Sor}, the created KK-modes does not simply escape away from 
the brane to the bulk in the high-energy region $\epsilon_*\gtrsim1$. 
They could be turned back to the brane by the curvature 
scattering of the anti-de Sitter bulk, leading to the significant 
enhancement or the modulation of the GWs on brane.

In order to investigate the degree of this effect precisely, 
the boundary condition imposed on the regulator brane might be 
inadequate. Instead of using the Neumann condition (\ref{eq:Neumann}),  
a suitable choice of the boundary condition such as the 
non-reflecting boundary condition should be considered. 
Further, to impose a realistic boundary condition, 
the Poincar\'e coordinate, which covers the wider region of the 
anti-de Sitter spacetime than the Gaussian normal coordinate,  
would be crucial in our numerical calculation. The implementation of these 
technical points is straightforward and the analysis is now in progress. 
We will report the results in a separate paper\cite{TH}.

\begin{ack}
We would like to thank R. Maartens for bringing closely related
work by R. Easther, D. Langlois, R. Maartens and D. Wands \cite{ELMW}
to our attention. 
K.K. acknowledges the supported of Japan Society for Promotion of 
Science(JSPS) Research Fellowships. A.T is supported 
by a Grant-in-Aid for Scientific Research from the JSPS(No.14740157).
\end{ack}


\end{document}